\documentstyle[twocolumn]{article}

\newdimen\headerboxheight
\newdimen\betweenumberspace          
\newdimen\aftertext                  
\newdimen\headlineindent             
\newcommand{\be}{\begin{equation}}
\newcommand{\ee}{\end{equation}}
\newcommand{\ra}{\rightarrow}

\newcommand{\beq}{\begin{equation}}
\newcommand{\eeq}{\end{equation}}

\begin{document}
\input{psfig}

\title{\bf On the dispersive two-photon $K_L\rightarrow \mu^+ \mu^-$ amplitude}

\author{J.O. Eeg\\Department of Physics, University of Oslo, N-0316 Oslo, Norway
(e-mail: j.o.eeg@fys.uio.no)
\and K. Kumeri\v{c}ki\\Department of Physics, Faculty of Science, 
      University of Zagreb,\\
    POB 162, HR-10001 Zagreb, Croatia (e-mail: kkumer@phy.hr)
\and I. Picek\\Department of Physics, Faculty of Science, 
      University of Zagreb,\\ 
    POB 162, HR-10001 Zagreb, Croatia (e-mail: picek@phy.hr)}

\date{August 1997 (revised and corrected version)}

\maketitle

\begin{flushleft}
Preprint BI - TP 96/08, Oslo-TP-2-96 and ZTF - 96/03\\
hep-ph/9605337
\end{flushleft}

\begin{abstract}
 We present a full account of the two-loop electroweak, two-photon mediated
short-distance dispersive $K_L \rightarrow\mu^+\mu^-$ decay amplitude.
QCD corrections change the sign of this amplitude and reduce it by
an order of magnitude.
Thus, the QCD-corrected two-loop amplitude represents
only a small fraction (with the central value of  5 \%) of the 
one-loop weak short-distance 
contribution, and has the same sign.
In combination with a recent measurement, the 
standard-model prediction of the short-distance
amplitude, completed in this paper,
provides a constraint on the otherwise uncertain
 long-distance dispersive amplitude.
\end{abstract}
 
\vskip0.5cm\hrule\vskip3ptplus12pt\null

\section{Introduction}

Even before it was measured, the $K_L \rightarrow\mu^+\mu^-$
decay had provided valuable insight into the understanding of
weak interactions. The non-observation of the
$K_L \rightarrow\mu^+\mu^-$ decay at a rate  comparable with that of
$K^{+}\rightarrow \mu^{+}\nu_{\mu}$ showed the importance of the
GIM mechanism \cite{GIM}: the invention of the charmed quark made
possible the
necessary suppression of the amplitude. 
Now, equipped with the results of the new measurements 
and in view of the forthcoming data, we take  this important amplitude
under scrutiny.

The amplitudes in a free-quark  calculation \cite{GL74}
(Fig.~1a and Fig.~1b)
represented by one-loop (1L) W-box and Z-exchange
diagrams, respectively, exhibited  a fortuitous 
cancellation of  the leading-order contributions.
Therefore, as shown by Voloshin and Shabalin \cite{VS76}, one was
addressed
to consider the two-loop (2L) 
diagrams corresponding to Fig.~1c
as a potentially important {\em light-quark} contribution.

\begin{figure}
\centerline{\psfig{figure=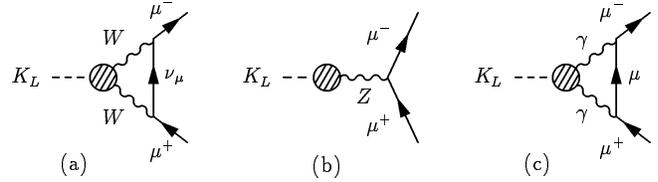,width=8.5cm}}
\caption{\footnotesize Possible mechanisms for $K_{L}\rightarrow\mu^{+}\mu^{-}$}
\end{figure}
 
The contributions shown in Fig.~1c  were brought to attention
by the measurements \cite{E791,KEK} which indicated that the {\em
absorptive}
part of the diagram in Fig.~1c dominated the rate of the
$K_L \rightarrow\mu^+\mu^-$ decay.
Namely, normalizing the amplitudes to the branching ratio 
\be
B(K_L \rightarrow\mu^+\mu^-)=|\mbox{Re} {\cal A}|^2 + |\mbox{Im} {\cal A}|^2
\;,
\label{bri}
\ee
and comparing it with the most recent BNL measurement \cite{E791}
\be
B(K_L \rightarrow\mu^+\mu^-)= (6.86 \pm 0.37)\times 10^{-9} \; ,
\label{branch}
\ee
exhibits the  saturation by the absorptive (Im${\cal A}$) part.
It completely dominates the $K_L \rightarrow\gamma\gamma
\rightarrow\mu^+\mu^- $ contribution, giving the so-called unitarity
bound \cite{unit}
\be
B_{\rm abs}=(6.8 \pm 0.3)\times 10^{-9} \; ,
\label{babs}
\ee
corresponding to Im${\cal A} = 8.25 \times 10^{-5}$.
Comparing the measurement (\ref{branch}) with the unitarity bound (\ref{babs}),
 there is  room for a total Re${\cal A}$ of order
$2 \times 10^{-5}$.
Thus, the total {\em real} part of the amplitude, being the sum of
short-distance (SD) and long-distance (LD) dispersive
contributions,
\be
\mbox{Re}{\cal A} ={\cal A}_{\rm SD} + {\cal A}_{\rm LD}  \;,
\label{rea}
\ee
 must be relatively small compared with the absorptive part of the
amplitude, as illustrated in Fig.~2. Such a small total dispersive
amplitude can be realized either when the SD and LD parts are both
small (Fig.~2a) or by partial cancellation between these two 
parts (Fig.~2b).
Notably, the opposite sign of SD and LD contributions (as favoured
by some calculations) leaves more space for an additional SD contribution.
If the SD amplitude is found to be small, there is no
room for a large LD dispersive amplitude  ${\cal A}_{\rm LD}$.
This leads us to reconsider previous SD calculations \cite{VS76,GMV}
in the next section.

\begin{figure}
\centerline{\psfig{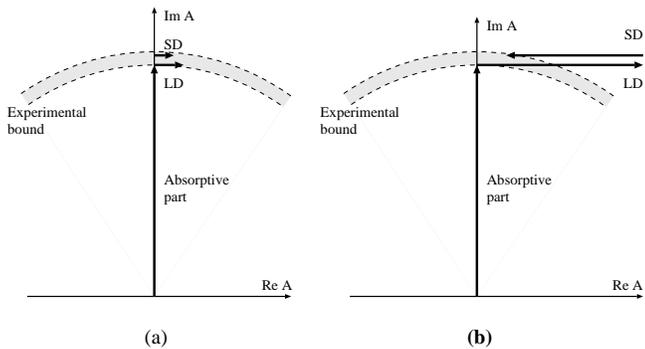}}
\caption{\footnotesize Schematic Argand diagram of the possible interplay of the 
         amplitudes under consideration}
\end{figure}

Frequently, the SD part has been identified as the
weak contribution  represented by the one-loop 
W-box and Z-exchange diagrams of Figs. 1a and 1b. This one-loop
contribution ${\cal A}_{1L} = {\cal A}_{\rm Fig.1a} + {\cal A}_{\rm Fig.1b}$  is
dominated by the $t$-quark in the loop 
(proportional to the small KM-factor $\lambda_{t}$), and the inclusion
of  QCD
 corrections  \cite{NSVZ,Buras} does
not change this amplitude essentially.
In the present paper we stress that the diagram of Fig.~1c
($\sim \alpha_{\rm em}^2 G_{\rm F}$) leads to the same
SD operator as that of preceding two diagrams (proportional
to $G_{\rm F}^2$). As already pointed out in 
\cite{VS76,SV79},
 the corresponding two-loop diagrams with two intermediate 
virtual photons have a short-distance
part ${\cal A}_{\rm 2L}$ (contained in 
${\cal A}_{\rm Fig.1c} = {\cal A}_{\rm LD} + {\cal A}_{\rm 2L}$) picking up 
a potentially sizable contribution 
from relatively high-momentum photons. The total SD amplitude is
\begin{displaymath}
        {\cal A}_{\rm SD}
={\cal A}_{\rm 1L} +  {\cal A}_{\rm 2L}\;.
\end{displaymath}
By exploring the contribution from Fig.~1c leading to the 
${\cal A}_{\rm 2L}$ amplitude, we isolate the 
strongly model-dependent LD dispersive piece.
Section 2 is devoted to the calculation of the dispersive two-loop
SD amplitude ${\cal A}_{\rm 2L}$. In section 3 we  conclude that
this amplitude
enables us to predict the possible range of the 
LD dispersive amplitude ${\cal A}_{\rm LD}$, the knowledge of which
has been urged by  studies of the related rare kaon decays \cite{DoG}.

\section{Dispersive two-loop SD contribution}

A complete treatment of the two-loop amplitude considered here
is a missing piece in the literature. 
There is  an enlightening feature of the diagram
in Fig.~1c:   the loop-momentum of the photon
in Fig.~1c  enables us to
control the distinction between the LD and SD contributions from this
diagram.  We approach this problem of separating the
two contributions by studying the SD piece, defined by the 
photon momenta above some infrared cut-off of the order
of some hadronic scale $\Lambda$. A sensible SD amplitude
should have a mild dependence on the choice of the particular
value of $\Lambda$. We
 calculate the (two-loop) {\em quark process}
\begin{equation}
     s\bar{d}\rightarrow\gamma\gamma\rightarrow\mu\bar{\mu}\;,
\label{2loop}
\end{equation}
for which we obtain a result proportional to
the left-handed quark current for the $s \rightarrow d$ transition.
\begin{figure}
\centerline{\psfig{figure=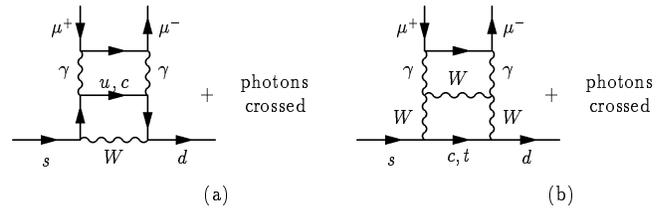,width=8.5cm}}
\caption{\footnotesize The dominant contributions to the $s\rightarrow d\gamma\gamma$
    induced 2L diagrams: $A1$ for the ($c, u$) quarks in the loop (a);
    $A3$ for the ($t, c$) quarks in the loop (b)}
\end{figure}
We present the main results of the calculation of the
full set of 44 electroweak (EW) two-loop diagrams in the 't Hooft-Feynman
gauge. It is convenient to distinguish between three sets of diagrams,
depending on one-particle irreducible subloops -- the $A$-diagrams
given by $s\rightarrow d\gamma\gamma$ transitions (of the type
shown in Fig.~3), the  $B$-diagrams
given by  the $s\rightarrow d\gamma$ transition (illustrated in
Fig.~4) and the $C$-diagrams given by the
non-diagonal $s\rightarrow d$ transition (shown in Fig.~5).
\begin{figure}
\centerline{\psfig{figure=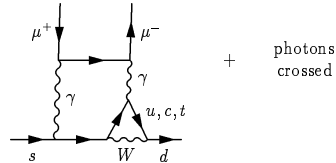,height=60pt}}
\caption{\footnotesize The dominant $B1$ contribution to the  $s\rightarrow d\gamma$
  induced 2L diagrams}
\end{figure}
\begin{figure}
\centerline{\psfig{figure=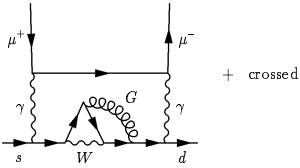,height=60pt}}
\caption{\footnotesize A genuine QCD contribution to $s\rightarrow d\mu^{+}\mu^{-}$,
 induced by the $s\rightarrow d$ self-penguin transition}
\end{figure}
We stress  that the $s\rightarrow d
\gamma\gamma$ electroweak insertions  are finite,
 whereas the divergent 
$s\rightarrow d\gamma$ 
and $s\rightarrow d$ insertions 
require a proper regularization. For the external light quarks
at hand, we have used the on-shell subtraction in the limit of vanishing
external 4-momenta. 
The structure for $C$-diagrams corresponds to the $s\rightarrow d$ amplitude
 regularized
 to be zero at the mass shells of  the $s$- and $d$-quarks \cite{Shab}, in 
the limit 
 $m_{s,d} \ra 0$, in which we work.
 
 After regularization,
the effective  $s\rightarrow d \gamma\gamma$ ($A$-tran\-si\-ti\-on),
$s\rightarrow d\gamma$ ($B$-transition)
and $s\rightarrow d$ ($C$-transition) have the structures
\begin{eqnarray}
{A:} \; \;
  \epsilon^{\mu \nu \sigma \rho} k_\sigma \bar{d}\gamma_{\rho} L s
\;,\;\;\; &&
{B:} \; \; (g^{\mu \rho} k^2 -k^\mu k^\rho) \bar{d}\gamma_{\rho} L s
\;, \nonumber \\
{C:} \; \; \bar{d} (\gamma \cdot k)^3 L s \;,\qquad && 
\label{struct}
\end{eqnarray}
 where $k$ is the photon-loop momentum (which for $B$- and $C$-diagrams
coincides with the $s$- or $d$-quark momentum inside the loop).
 After regularization, all three types
 of diagrams are 
internally gauge invariant with respect to QED when diagrams with crossed 
photons are added.  Other structures, besides those in (\ref{struct}),
are present for 
($A$) $s\rightarrow d \gamma\gamma$ and ($B$) $s\rightarrow d\gamma$ diagrams,
 but  do not contribute to the two-loop quark process (\ref{2loop}) when
 diagrams with crossed and uncrossed photons are summed.
 
The two-loop amplitude 
resulting from  the
$A$, $B$ and $C$ subloops in (\ref{struct}) acquires the form
\begin{equation}
{\cal{M}}_{2\gamma}^{q} = \frac{i G_{\rm F}}{2\sqrt{2}}\frac{\alpha^2}{\pi^2}
           \lambda_q \left\{ A_q + B_q + C_q \right\} 
    (\bar{d} \gamma^{\beta}Ls) (\bar{u}\gamma_{\beta}\gamma_5 v) \;,
\label{2loop2}
\end{equation}
which is proportional to the same operator as that appearing 
in the one-loop amplitude 
\cite{NSVZ,Buras}.
Summing over the quark flavours ($q = u,c,t$) in the loop gives us a general
amplitude as
\begin{eqnarray}
{\cal M} (s\bar{d}\rightarrow\mu\bar{\mu})&=&
 \sum_{q} \lambda_{q}{\cal M}^{q} \nonumber \\
  &=& \lambda_{u}({\cal M}^{u} 
  - {\cal M}^{c}) + \lambda_{t}({\cal M}^{t} -  {\cal M}^{c}) \nonumber \\
 &=& -\lambda_u {\cal M}^{(c,u)} +
  \lambda_t {\cal M}^{(t,c)} \;,
\label{acal}
\end{eqnarray}
explicitly exposing the GIM mechanism (the $\lambda_q$'s  are the
appropriate KM factors).
After embedding the  $s\bar{d}$ ($d\bar{s}$) in the meson $\bar{K}^0$
($K^0$), the physical CP-conserving  amplitude 
 takes the form
\begin{equation}
{\cal A}(K_L\rightarrow\mu\bar{\mu})^{\mbox{\scriptsize CP-cons}}
= -\lambda_u {\cal A}^{(c,u)}+\mbox{Re}\lambda_t {\cal
A}^{(t,c)} \;.
\label{cpc}
\end{equation}
 
For the light quarks ($q= c, u$), diagrams $A1$ (Fig.~3a)
 and $B1$ (Fig.~4) dominate completely (and are therefore under
scrutiny in Table 1), the other diagrams being suppressed
 by an extra factor $m_c^2/M_W^2$ after the GIM mechanism has been taken
 into account.
For the heavy quark ($t$) in the loop, such a suppression is of course absent,
 and we a priori have  to consider all diagrams.
It turns out that then the largest contribution among $A$-diagrams is $A3$ 
(Fig.~3b), and among $B$-diagrams the largest is again $B1$. 
Both in the light-  and heavy-quark cases there
are also the contributions from the non-diagonal $s \rightarrow d$ 
self-energy ($C$-diagrams).
 Being negligible in the
pure electroweak case (suppressed by $m_c^2/M_W^2$ for  light
 quarks after GIM),
 the off-diagonal self-energy contribution becomes potentially
unsuppressed ($\sim \alpha_s \ln m_c$) when perturbative QCD is
switched on \cite{SP} (Fig.~5). 

\subsection{Pure electroweak results}

Let us first display the {\em pure electroweak} (EW) results in
order to keep contact with the early calculation by Voloshin and 
Shabalin \cite{VS76}.
We have calculated all the contributions numerically, the results of the 
dominating ones being presented in Table 1. In addition,
the analytical expressions can be obtained in the {\em light}-quark
($u, c$) case.
Let us display the analytic forms for the leading $A1$ and $B1$ amplitudes 
which reproduce those obtained  previously \cite{VS76}. 
Our calculation shows that, for $m^{2} \ll M_{W}^{2}$, the leading logarithmic
 (LL) contribution in the pure electroweak case is
\begin{equation}
  A1_{\rm LL} = -\frac{2}{3}\left[ \ln\frac{M_{W}^{2}}{\Lambda^{2}} -
  2\ln\frac{m^{2}}{\Lambda^{2}}\right]\; ,
\label{A1LL}
\end{equation}
where $\Lambda$ is the infrared cut-off, defined above. In this way
we avoid integrals over low photon momenta, which correspond to some
LD contributions. 
 For the amplitude $B1$, we obtain the
following LL result for the single
quark loop (for $m^{2} \ll M_{W}^{2}$):
\begin{eqnarray}
B1_{\rm LL} &=& -\frac{4}{9}\left[\frac{1}{2}(\ln\frac{M_{W}^{2}}{\Lambda^{2}})^2 -
 \frac{1}{2} (\ln\frac{m^{2}}{\Lambda^{2}})^2\right. \nonumber \\
    &&+\left.\frac{5}{6}\ln\frac{M_{W}^{2}}{m^{2}}
    -\frac{5}{6}\ln\frac{m^{2}}{\Lambda^{2}}
     \right] \;.
\label{B1LL}
\end{eqnarray}
Taken at face value, the expressions (\ref{A1LL}) and  (\ref{B1LL})
 are the result for the $c$-quark case ($m= m_c$).
 The corresponding $u$-quark contribution
is obtained by the replacement 
$m_u \rightarrow \Lambda$.
 These results conform to \cite{VS76} after the GIM mechanism
has been taken into account.

As a new contribution to the existing literature, we have also performed
the 2L calculation of the
electroweak diagrams for the {\em heavy} quarks ($q = t, c$) in 
the loop. 
In this case, the dominant contributions are $A3$ (Fig.~3b) and $B1$ 
(Fig.~4). However, these are associated with the small KM factor $\lambda_t$
and  are therefore suppressed. Table~1 displays only
these dominant amplitudes and the total amplitudes, a full account being
relegated to a more detailed publication \cite{EKPprep}.
This table also illustrates a mild sensitivity
of the dominant light-quark electroweak amplitudes $A1$ and $B1$
to the IR cut-off $\Lambda$.
As we have also displayed the total amplitude, this table illustrates to
what degree the indicated contributions
are dominant within the full set of the pure EW diagrams.
The agreement between the numerical ($A1$ and $B1$) and the analytical 
LL results (\ref{A1LL}) and (\ref{B1LL}), after performing the GIM
procedure,
is explicated by the corresponding rows of Table 2.
The last row of Table 1, normalized to the measured amplitude,
shows the largeness of the net EW contribution.
We observe that such a large pure electroweak 2L contribution would have
decreased the one loop  amplitude \cite{Buras} substantially. 

\begin{table*}
\caption{\footnotesize The pure electroweak light-quark ($c, u$) and heavy-quark ($t, c$) 
2-loop results. 
The input values are $m_{t}$=173~GeV, 
  $m_{c}=1.5$~GeV and $m_{u}$ replaced by the
  IR cut-off $\Lambda$ of 0.7 or 0.9~GeV
(corresponding to the range considered in  (\protect\ref{result})).
The values in the last row are obtained by multiplying by
$\lambda_{u}=0.215$ or $\lambda_{t} \simeq 2\times10^{-4}$, and should be
compared with $|{\cal A}_{\rm expt}|\simeq|$Im${\cal A}|=
\protect\sqrt{B_{\rm abs}}=8.25\times 10^{-5}$.
  The one-loop 
 (1L) SD contribution
corresponds to a Re${\cal A}$ (see (\protect\ref{BuBu}))
of the order $-3.5 \times 10^{-5}$}
\begin{center}
\renewcommand{\arraystretch}{1.1}
\begin{tabular}{crrcrr}
\hline Dominant
 &\multicolumn{2}{c}{Pure EW} & Dominant&\multicolumn{2}{c}{Pure EW}\\
$(c,u)$ diagram & $\Lambda=0.7$ & $\Lambda=0.9$&
$(t,c)$ diagram& $\Lambda=0.7$ & $\Lambda=0.9$ \\ \hline
      $A1$ & 1.86 & 1.22 & $A3$     & 23.8    & 22.7    \\ 
$A$ total & 1.90 & 1.25 &  $A$ total  & 27.1    & 26.4   \\ \hline
      $B1$ & 1.70 & 1.04 &  $B1$      & 20.6    & 19.1   \\ 
$B$ total & 1.70 & 1.03 &  $B$ total  & 15.6    & 14.2   \\ \hline
 Total   & 3.59 & 2.28 &  Total     & 42.7    & 40.7   \\ \hline
 Re ${\cal A} / 10^{-5}$&1.55 &0.98& Re ${\cal A} / 10^{-5}$&
 0.017& 0.016  \\ \hline
\end{tabular}
\end{center}
\label{tbl1}
\end{table*}
 
\subsection{QCD corrections}

There are some subtleties in performing  QCD corrections
to the two-loop diagrams considered. Although the gluon corrections
pertain to the quark loop, the highly off-shell photons closing the other
(quark-lepton) loop control the SD regime of the two-loop amplitude
as a whole.
In general, there is up to one log per loop, as exemplified by the
$B1$-term in (\ref{B1LL}) 
 related to Fig.~4.

There is a suitable prescription introduced in Refs. \cite{NSVZ,SVZemp}
and applied by other groups \cite{Oth,PRDjoe,IPi} for handling 
the leading
QCD corrections. Using this prescription, one can write the amplitude
as an integral over virtual quark loop momenta. In the
problem considered, we have to
decode the 2-loop momentum flow in order to extract
the leading logarithmic structure,
which we then sum using the renormalization-group technique. 
Thereby, we refer to the building blocks considered 
previously -- the electromagnetic penguin of Ref. \cite{EP88} 
(now appearing in the
$B1$ amplitude), the QCD corrections to the quark-loop of Fig.~3a
\cite{ENP90} and to a very recent treatment of the self-penguin \cite{BeE}.
Let us present this in more detail. 

We start by demonstrating the QCD corrections to the $c$- and $u$-quark loops
of $A$-diagrams in Table 2. One first hunts the leading log 
 which should
correspond to the $A1$-term in (\ref{A1LL}).
This result  can be understood from the result 
of the previous  
$ s \bar{d} \rightarrow \gamma \gamma$
calculation \cite{ENP90}, which consisted of two terms dominated at the
scales $M_W$ and $m$, respectively. Moreover, these two terms had the relative
weights 1 and $-2~$, respectively.
When this $ s \bar{d} \rightarrow \gamma \gamma$
amplitude is inserted in to the two-loop diagram for 
$ s \bar{d} \rightarrow \mu \bar{\mu}$, we gain one logarithm.
 Since the two terms in (\ref{A1LL})
stem from the loop integrals dominated by the momenta at $M_{W}^{2}$
and $m^{2}$, respectively, the QCD-corrected amplitude acquires the form
\begin{equation}
  A1_{\rm LL}^{\rm QCD} = -\frac{2}{3}\eta_{1}(M_{W}^{2})
    \ln\frac{M_{W}^{2}}{\Lambda^{2}}+ 
      \frac{4}{3}\eta_{1}(m^{2})\ln\frac{m^2}{\Lambda^{2}} \; ,
\label{A1LLQCD}
\end{equation}
which in principle agrees 
with \cite{VS76} and disagrees with \cite{GMV}.
Here, the QCD coefficient $\eta_1$ reflects the colour-singlet nature of
the photonic part of the diagram, and is given by
\begin{equation}
   \eta_{1}(q^{2}) = 2 c_{+}(q^{2}) - c_{-}(q^{2})\; ,
\label{sing}
\end{equation}
where $c_{\pm}$ are the Wilson coefficients
 of the 4-quark operators ${\cal O}_{\pm}$ in the effective
$\Delta S = 1$ Lagrangian of Ref. \cite{GLAM}. In the leading
logarithmic approximation they are given by
\begin{equation}
   c_{\pm}(q^{2})=\left[\frac{\alpha_{s}(q^{2})}{\alpha_{s}(M_{W}^2)}
    \right]^{{a_{\pm}}/{b}} \;,
\label{cQCD}
\end{equation}
where  $a_{+}=-2$ and $a_{-}=4$ are the anomalous dimensions
and $b=11-2 N_{f}/3$, $N_f$ being the number of active flavours.
In contradistinction to the numerically favourable and stable colour-octet
factor $\eta_{8} = (c_{+}~+~c_{-})/2$, the singlet coefficient
(\ref{sing}) is rather sensitive to the choice of  $\Lambda_{\rm QCD}$,
with a notable switch of the sign \cite{SVZemp,PRDjoe}
 for $q^{2}$ at the scale of a few GeV$^{2}$.
 Combining the
$u$- and  $c$-quark contributions by taking into account the GIM
mechanism (see (\ref{acal})), only the second term in
(\ref{A1LLQCD}) survives. 

The $B1$ amplitude in (\ref{B1LL}) can be understood in terms of the
 electromagnetic 
penguin subloop, which is, within the LL expansion, proportional to
\begin{displaymath}
\ln (\frac{M_W^2}{m^2}) - \frac{5}{6} \quad  \mbox{for}
 \quad  k^2<m^2<M_W^2 \;,
\end{displaymath}
\begin{equation}
\ln (\frac{M_W^2}{k^2}) + \frac{5}{6} \quad  \mbox{for}
\quad  m^2<k^2<M_W^2 \;,
\label{empeng}
\end{equation}
where $k$ is  the momentum of virtual photons. Inserting this subloop
 into the next loop, we gain one logarithm (in particular, 
ln $\rightarrow$ ln$^2/2$).
Hence the $log^2$ form in the second term in (\ref{B1LL}),
which leads to  the QCD-corrected amplitude
expressed in an integral form as
\begin{eqnarray}
B1_{\rm LL}^{\rm QCD}&=& -\frac{4}{9}\left[\int_{m^2}^{M_{W}^{2}}
\frac{dp^{2}}{p^{2}} \eta_{1}(p^{2}) \left( \ln\frac{p^{2}}{\Lambda^{2}} + 
\frac{5}{6} \right)\right. \nonumber \\
 &&-\left.\frac{5}{6}\eta_{1}(m^{2})\ln\frac{m^2}{\Lambda^{2}} 
  \right] \;.
\label{B1LLQCD}
\end{eqnarray}
Again, the expressions (\ref{A1LLQCD}) and (\ref{B1LLQCD}) apply
directly to the $c$-quark contribution, the $u$-quark contribution
being obtained by making the replacement $m\rightarrow\Lambda$. 
This means that when taking into account the GIM mechanism, the integral
in (\ref{B1LLQCD})
will run from $\Lambda^2$ to $m_c^2$.
The net result of the QCD dressing is similar to that for the
$A1$ diagram: a suppressed amplitude with a change of sign. 

The $C$-contribution stemming from the QCD-induced self-penguin 
($SP$) in Fig.~5. might also be interesting.
As opposed to $A1$ and $B1$ contributions it is not suppressed
 by the colour singlet factor $\eta_1$, but contains the numerically 
favourable colour octet factor $\eta_8$. It is, however, suppressed
 by $\alpha_{\rm s}/\pi$.
For the $m=m_{c}$ case, we obtain to all orders in QCD
\begin{eqnarray}
C_{\rm LL}^{\rm QCD}&=& \frac{7}{162}\left[\int_{m^2}^{M_{W}^{2}}
\frac{dp^{2}}{p^{2}}
 \rho(p^2) \left[ \frac{1}{2}(\ln\frac{p^{2}}{\Lambda^{2}})^2
 \right.\right. \nonumber \\
 &+&\left.\left.
  (\frac{5}{6}+ \frac{25}{21})\ln\frac{p^{2}}{\Lambda^{2}} \right] 
 -\frac{5}{6} \rho(m^{2}) \frac{1}{2}(\ln\frac{m^2}{\Lambda^{2}})^2 
 \right] \; ,
\label{CSP}
\end{eqnarray}
where $\rho(p^2) = \eta_8(p^2) \alpha_{\rm s}(p^2)/\pi$. 
In addition, 7/162 is an 
overall loop factor, and the terms 5/6
have the same origin as in  (\ref{empeng}) and (\ref{B1LLQCD}).
 The $u$-quark 
contribution is again obtained by making 
the replacement $m \rightarrow \Lambda$.

The light-quark approximation ($m^2 \ll M_W^2$) is used in (\ref{CSP}).
For an arbitrary quark mass, needed to treat the heavy top in the loop,
 the calculations are much more difficult \cite{BeE}.
 We have done an estimate and found
that the top-quark contribution is roughly 10 $\%$ of the charm-quark
contribution 
 (taking into account that $\alpha_{\rm s}$ at $m_t$
 is about 1/3 of $\alpha_{\rm s}$ at $m_c$.).

Table 2 displays a  detailed structure of the dominant amplitudes
from Table 1, before and after applying the GIM mechanism: 
the first, the second, and the third block of the table
display the $A$, $B$, and $C$ contributions, respectively.
In the third block, $C_{\rm LL}^{\rm SP}$ and $C_{\rm LL}^{\rm QCD}$ 
refer to the bare
and dressed self-penguin
contributions, respectively, 
whereas $C$ refers to the negligible pure electroweak 
(EW) contribution.
Therefore, $C_{\rm LL}^{\rm SP}$ is different from $C$.
As a curiosity, we have found that the latter has a peculiar GIM
cancellation: there is an exact cancellation between the
$c$-quark contribution for $m_c \rightarrow 0$ and the $t$-quark contribution 
for $m_t \rightarrow \infty$.
 As a result,
$C$ is not so GIM-relaxed as expected for a heavy-quark case $(t, c)$.

\begin{table*}
\caption{\footnotesize The anatomy of QCD corrections:
the exact EW 2-loop calculation is compared with the LL values
and with the RGE summed LL QCD corrections.
The input values are the same as in Table 1, with the IR cut-off specified
at $\Lambda$=0.83~GeV and with $\alpha_{\rm s}(M_Z)=0.118$
\protect\cite{PDG}. 
Correspondence with the empirical value can be made
using the conversion factors provided by the last row of Table 1}
\begin{center}
\renewcommand{\arraystretch}{1.4}
\begin{tabular}{crrrcrrr} \hline
  &\multicolumn{3}{c}{EW + SD QCD}& & \multicolumn{3}{c}{EW + SD QCD} \\
  & $c$-loop & $u$-loop &GIM $(c-u)$ &  & $t$-loop &
          $c$-loop & GIM $(t-c)$  \\ \hline
$A1$ &-2.14 &-3.57 &1.42& $A3$ & -17.4&-40.4&23.0   \\ 
$A1_{\rm LL}$&-4.52 &-6.09&1.58&$A3_{\rm LL}$&-22.3&-18.3&-4.0 \\ 
$A1_{\rm LL}^{\rm QCD}$& -6.19 &-6.09&-0.10&
$A3_{\rm LL}^{\rm QCD}$&-22.3&-18.3&-4.0\\ \hline

$B1$ &-20.4&-21.6&1.24&$B1$&-0.8&-20.4&19.6    \\ 
$B1_{\rm LL}$&-20.8&-22.0&1.19&$B1_{\rm LL}$&-2.0&-20.8&18.8  \\ 
$B1_{\rm LL}^{\rm QCD}$&-14.9&-14.7&-0.23&
$B1_{\rm LL}^{\rm QCD}$& -2.0&-14.9&12.9 \\ \hline

$C$ &-0.61&-0.61&$3 \times  10^{-3}$& $C$ &-0.52 &-0.61& 0.09 \\ 
$C_{\rm LL}^{\rm SP}$&0.46& 0.47& -0.01 &$|C_{\rm LL}^{\rm SP}|$&  
$< 0.2  $ &0.47  &$ < 0.7 $ \\ 
$C_{\rm LL}^{\rm QCD}$& 0.47 &  0.48  & -0.01 &
$|C_{\rm LL}^{\rm QCD}|$&  $ < 0.2 $&  0.47 &$ < 0.7  $  \\ \hline
\end{tabular}
\end{center}
\label{tbl2}
\end{table*}

\section{Conclusions}

In this paper we have focused on the 2-loop (2L)
 contributions,
leading to the typical SD local operator for the
$s\bar{d}\rightarrow\gamma\gamma \rightarrow\mu^+\mu^- $
quark transition but also having a LD (soft-photon) range.
Our approach starts from the SD side, 
whereby an infrared (IR) cut-off of virtual photons sets in.
We contrast the SD contribution with the complementary LD ones,
 which have to be calculated using other methods, and are rather model 
dependent at the present stage. The numerically important 
2L {\em pure electroweak} SD contributions
 are due to the light ($u, c$)
quarks in the loop. 
Besides completing the previous calculation for light quarks,
we have also considered the 2L diagrams including the
heavy ($c, t$) quarks.
A large number of electroweak diagrams may compensate for
a small CKM factor,  and one might expect non-negligible effects.
However, the actual calculation shows that the various amplitudes
 have different signs, and taking into account the smallness of
$\lambda_t$, the heavy-quark contribution is negligible.

Next, we have shown the
importance of the SD QCD corrections for the 2L diagrams, summarized
in Fig.~6.
\begin{figure}
\centerline{\psfig{figure=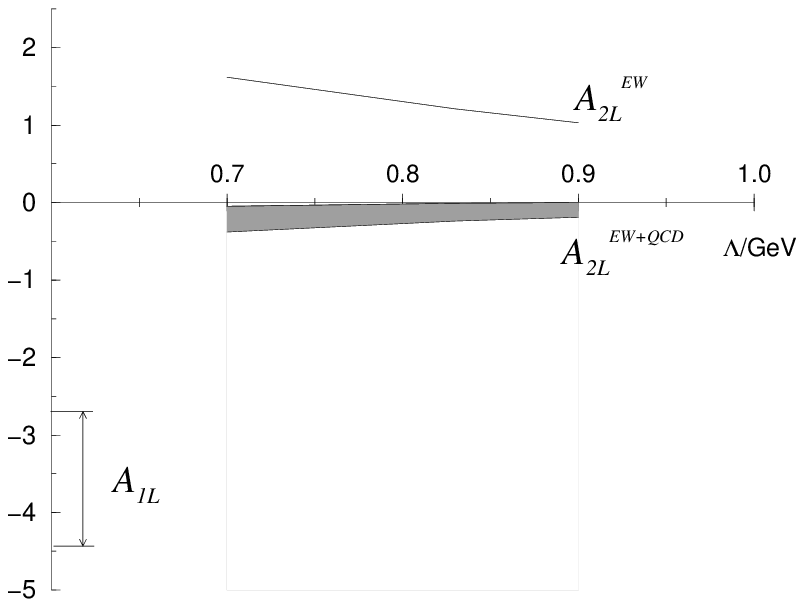}}
\caption{\footnotesize QCD does three things to the EW 2L amplitude:\protect\newline
  \hspace*{5mm} i) smoothens the $\Lambda$-dependence (making the SD
   extraction better defined),\protect\newline
\hspace*{4mm} ii) changes the sign (making it coherent to 
   the 1L amplitude), and \protect\newline
\hspace*{4mm} iii) suppresses it to large extent}
\end{figure}
Inclusion of these QCD corrections  appears to be  subtle and
more dramatic than it was the case for the 1L diagrams. 
Two decades ago there was a controversy concerning QCD corrections 
to these 1L diagrams. Ref. \cite{NSVZ}
resolved it by an adequate treatment of the loop integrals.
Our results for SD corrections to the 2L diagrams are shown in Table~2.
The short-distance QCD corrections suppress the part of the
SD $2\gamma$ amplitude  which is electroweakly dominant
before inclusion of QCD corrections.
 The basic reason for this is the
behaviour of
the $\eta_1 = (2c_{+}-c_{-})$ QCD coefficient. 
In particular, the $A1$ and $B1$ amplitudes are suppressed
 to a
large extent, and do not anymore interfere destructively with
 the 1L SD amplitude
of Ref. \cite{Buras}. Without this suppression, the scenario
of Fig.~2a would appear as a more likely one.

We should stress that in the treatment of the 2L amplitude we have performed
QCD corrections in the leading logarithmic approximation by using 
(\ref{cQCD}), while the 1L amplitude was treated in the next 
to leading (NLO) approximation in \cite{Buras}.

To summarize, we have found a modest light-quark 2L contribution stemming from 
intermediate virtual photons having relatively high momentum. 
Introducing the error bars corresponding to $\Lambda$ in the 
range 0.7--0.9~GeV, and a more essential one from empirical uncertainty
in $\alpha_{s}$ (corresponding to $\Lambda^{(5)}_{\rm QCD}$ in the
range 150--250~MeV), we obtain
\begin{equation}
   -0.38 \times 10^{-5} \leq {\cal A}_{\rm 2L}  \leq  -0.001\times 10^{-5} \; ,
\label{result}
\end{equation}
This has the same sign and, for central values, corresponds to 5 \% of
${\cal A}_{\rm 1L}$ \cite{Buras},
\begin{equation}
   -4.4\times 10^{-5} \leq {\cal A}_{\rm 1L}  \leq  -2.6\times 10^{-5} \; ,
\label{BuBu}
\end{equation}
where the uncertainty mainly reflects the poor knowledge of $\lambda_t$.
Although the 1L and 2L contributions are not treated on an equal footing
(NLO versus LL QCD corrections), 
this result still  enables us 
to estimate the size of ${\cal A}_{\rm LD}$ from (\ref{rea}). Referring to
our comments below (\ref{babs}), and allowing for a
$|{\mbox Re}{\cal A}| \leq 2.7\times 10^{-5}$, we find the following allowed
range for ${\cal A}_{\rm LD}$:
\begin{equation}
   -0.1\times 10^{-5}  \leq {\cal A}_{\rm LD} \leq 7.5\times 10^{-5} \; \, .
\label{LDamp}
\end{equation}
Thus, having a  dispersive LD part 
${\cal A}_{\rm LD}$ of the size comparable with the  
absorptive part \cite{DAm86}
is still not ruled out completely.

The two vector-meson dominance calculations for the LD amplitude
 considered as the referent calculations in Ref. \cite{E791}
have basically opposite signs:  
\[
  -2.9\times 10^{-5} \leq {\cal A}_{\rm LD} \leq 0.5\times 10^{-5}
  \quad \cite{BMS}\;,
\]
\begin{displaymath}
   0.27\times 10^{-5} \leq {\cal A}_{\rm LD} \leq 4.7\times 10^{-5}
  \quad \cite{KO}\;.
\end{displaymath}

On the basis of the inferred relative sign between 1L and
2L contributions, Ref. \cite{GMV}, attempted to 
discriminate between the two LD
calculations quoted above.
(They favoured Ref. \cite{BMS}, and disfavoured Ref. \cite{KO} as the one
ascribing opposite signs to SD and LD.)
In the last of their papers \cite{GMV} they even concluded that
the BNL measurements \cite{E791} were in conflict with the standard model.

We have found that these conclusions are doubtful, since they are
based on an erroneous SD extension to the LD momentum region.
In our opinion, Ref. \cite{GMV} 
misidentifies  what (according to the calculational
method employed) should be their SD amplitude
 ${\cal A}_{\rm 2L}$, with ${\cal A}_{\rm LD}$. 
 In our treatment (see section 2) we have avoided the forbidden 
low-momentum region by introducing the infrared cut-off $\Lambda$ of
the order of the $\rho$-mass. We have demonstrated that
there is a subtle QCD suppression of the originally
 quite sizable SD EW 2L
amplitude.
Therefore, a  real
$K \rightarrow\gamma\gamma \rightarrow\mu^+\mu^- $
amplitude of a considerable size given in (\ref{LDamp})
corresponding to low $\gamma$-momenta ($\sim \Lambda$ and below),
is still allowed.
This might be used as a consistency check for the methods of the
type employed in Refs. \cite{BMS,KO}.

Taking into account the difficulties inherent to the estimates of the LD 
amplitude, it is welcome to arrive at the constraint (\ref{LDamp}).
Accordingly, provided the sign of ${\cal A}_{\rm LD}$ are correctly given in
\cite{E791}, the BNL experiment combined with the standard-model calculation
tends to favour the result of Ref. \cite{KO}. In this way, the
 scenario of Fig.~2b
seems to be preferred by the standard model. Provided that the 
beyond-standard-model effects  are 
represented by the relatively small SD amplitudes, this
scenario hinders the possibility of recovering such effects 
in the $K_{L}\rightarrow\mu^{+}\mu^{-}$ decay.
The forthcoming data from $K_{L}\rightarrow\mu^{+}\mu^{-}$ measurements  
\cite{LP95} will further test the conclusions of the present paper.

\subsection*{Acknowledgement}
\footnotesize
Two of us (K.K. and I.P.)  gratefully acknowledge the partial support of the 
EU contract CI1*-CT91-0893 (HSMU) and the hospitality of the Physics
Department of the Bielefeld University. One of us (I.P.) would also like
to acknowledge the hospitality of the Department of Physics in Oslo, and
 to thank the Norwegian Research Council for a traveling grant.

\end{document}